\documentclass[12pt,oneside,A4paper,onecolumn]{paper}
\usepackage[]{fontenc}
\usepackage[dvips]{epsfig}
\usepackage{a4wide}
\usepackage{natbib}
\usepackage[rightcaption]{sidecap}
\usepackage{hyperref}

\newcommand{\BH}{BH}
\newcommand{\SMBH}{SMBH}

\newcommand{\apj}{ApJ}
\newcommand{\apjs}{ApJS}
\newcommand{\apjl}{ApJL}
\newcommand{\aap}{A{\&}A}

\newcommand{\mnras}{MNRAS}
\newcommand{\aj}{AJ}

\newcommand{\nar}{NewAR}

\begin{document}

\begin{center}
\textbf{{\Large{Seeking for the leading actor on the cosmic stage: Galaxies vs Supermassive Black Holes}}}
\end{center}
\vspace{0.5cm}
\noindent
{\bf Editors: A. Bongiorno$^{1,2}$, F. Shankar$^{3}$, F. Civano$^{4}$, I. Gavignaud$^{5}$, A. Georgakakis$^{1}$}\\

\begin{footnotesize}
\noindent
$^{(1)}$ Max-Planck-Institut f\"ur extraterrestrische Physik (MPE), Giessenbachstrasse 1, D-85748, Garching bei M\"unchen, Germany. \\
$^{(2)}$ Osservatorio Astronomico di Roma, via Frascati 33, 00040 Monteporzio Catone, Rome, Italy;\\
$^{(3)}$ GEPI, Observatoire de Paris, CNRS, Univ. Paris Diderot, 5 Place Jules Janssen, 92195 Meudon, France;\\ $^{(4)}$ Smithsonian Astrophysical Observatory, 60 Garden Street, MS 67, Cambridge, MA 02138 USA;\\
$^{(5)}$ Dept. Ciencias Fisicas, Facultad de Ingenieria, Universidad Andres Bello, Avda. Republica 252 Santiago, Chile; \\
\end{footnotesize}

\vspace{0.5cm}
A major development in  extragalactic astrophysics in recent years has
been the realization that Active Galactic Nuclei (AGN), which signpost
accretion  events onto supermassive  black holes  (SMBHs) \citep{Merloni2008,Shankar2009}, may  play a
fundamental    role    in    the    formation   and    evolution    of
galaxies.  Understanding the  physics that  drive the  growth  of SMBH
across  cosmic  time is  therefore  important  for  having a  complete
picture  of galaxy  formation.  The two processes of galaxy and BH evolution can  no longer  be
regarded as  separate, as was the  case until about 10  years ago, but
need to be studied in conjunction.

One of the first indications for a correlation between the formation
of galaxies and  the growth of SMBHs at their  centres is the striking
similarity  between the  redshift evolution  of the  accretion density
 and  the  star-formation  rate  density \citep{Merloni2004,Hopkins2006a} of the  Universe. Both quantities  show a
rapid increase from  the local Universe to $z\approx1$,  followed by a
broad plateau at $z\approx2-4$ and  a decline at $z\ge4$, although the
high redshift behavior of  X-ray AGN remains controversial \citep{Brusa2009hzlf}.

These similarities are more than a mere coincidence or a manifestation
of the fact that the Universe was overall more active in the past. 
Observational evidence now show that most, if not all, galaxies in the local universe posses a central SMBH at their center. Moreover, there is now emerging consensus from dynamical observations, that \SMBH s 
at the center of the local massive and bulge-dominated galaxies, 
are tightly correlated with the velocity dispersion and masses of their stellar hosts, with an intrinsic scatter of 
a factor of two or even less \citep{Magorrian1998,Ferrarese2000,Haring2004}. 
Such strong correlations  argue  for  a  physical
association  between  the   two  processes and suggest that \BH s must have evolved, or co-evolved,
with their stellar hosts at some point in their past.



Analytical  calculations identify AGN  feedback as the  process that can potentially  link 
\SMBH\ growth and star-formation \citep[e.g.][]{Silk1998,Fabian1999,King2003,Granato2004,Croton2006,Menci2006}.  
In this  picture, the energy released by the AGN is sufficient to either heat up or blow away the cold gas of galaxies, thereby  irreversibly  altering   their  evolution.   Observations  of
powerful  and/or nearby  active  \SMBH s have  recently started  finding
evidence for outflows, most likely  associated with the AGN, in either
the warm \citep[e.g.][]{Holt2011} or the  cold \citep[e.g.][]{Feruglio2010, Sturm2011} 
gas component of  their host galaxies. 


Despite the  increasing evidence  for the importance  of SMBH growth in
galaxy  formation, we have just  started exploring  the physical
processes at play.   The nature of AGN feedback, how  it is related to
the fueling mode of the \SMBH,  and ultimately what is its impact on kpc
or  even Mpc  scales are  still debated.  

To  shed  light on  these  issues from  both the  observational  and
theoretical  point of  views, we  initiated the  Special  Issue \textit{``Seeking for the leading actor on the cosmic stage: Galaxies vs Supermassive Black Holes''} (\href{http://www.hindawi.com/journals/aa/si/610485/}{\texttt{http://www.hindawi.com/journals/aa/si/610485/}})  which
collects  reviews and  new results,  contributed  by some  of the  key
researchers  in the field.   We made  an effort  to bring together  observers and
theoreticians  to  provide the community  with a comprehensive  state-of-the-art  overview  of  our  understanding  of \SMBH\ evolution across cosmic time. In this respect, the Special Issue deals with topics from the smallest to the largest scales including e.g. BH accretion mechanisms and properties, the  influence on their host galaxies, and the connection to the Dark Matter halos. We  hope  this Special Issue  will become  a  useful reference for all researchers in the field.

The Special Issue starts  with a paper from {\it S. Bianchi, R.  Maiolino,
\& G.  Risaliti}, who discuss the recent developments on the
AGN Unified Models. They review the standard Unified Model to then move to an updated Unification scenario that can better explain the complex phenomenology observed. This paper is followed by a series of papers which constitute a comprehensive  description of the  accretion  history  of  the  Universe. In particular, {\it F. Fiore,  S.  Puccetti,  \& S.   Mathur} present the z$>$3 X-ray sources number counts in the 0.5-2 keV band and make predictions for new missions;  {\it E. Treister \& C.  Urry} review what we  know about the energy output of AGN describing the most common way to isolate these sources at different wavebands. They also summarize the cosmic history of black hole accretion, i.e., when in the history of the Universe supermassive black holes were obtaining most of their mass.  {\it B.Kelly \& A.  Merloni} present the demographic of SMBHs in  the local Universe discussing advantages and disadvantages of the different methods for estimating the black hole mass function, and {\it A. Sesana} reviews the current understanding of massive BH formation and evolution along the cosmic history highlighting which future observations will help to shed light on the cosmic history of \SMBH s, paying particular attention to the upcoming gravitational wave window. 
A comprehensive discussion of current ideas of
\BH s fueling mechanisms is presented by {\it G.  Lodato},  while {\it T.  Johannsen} discusses possible new 
observational tests for confirming the existence of \BH\ at the centers of galaxies.   

However, \BH s
do  not   grow  only   via  gas  accretion   but  possibly   also  via
\BH-\BH\  mergers.  {\it M.  Dotti, A.   Sesana \& R.  Decarli} review  the
state-of-the-art of mergers as an additional channel of  \BH\ growth including the expected observational signatures of massive binaries.  In this context, {\it S.  Komossa}  presents the results of numerical relativity simulations which imply that after binary coalescence in a galaxy merger the newly formed single SMBH can receive kick velocities up to several 1000 km/s due to anisotropic emission of gravitational waves. Moreover, she presents the observational signatures of recoiling SMBHs and the properties of the first candidates which have emerged.

After that, we devoted part  of  the special  issue to our  current understanding of the connection  between \SMBH s
and  their host  galaxies  inferred from  a  variety of  observational
probes.   The  latest  results  on the  AGN/Starburst  connection  are
presented by {\it E.  Sani \& E.   Nardini} who studied a peculiar Ultraluminous Infrared Galaxy (IRAS 20551-4250) experiencing an intense starburst but hosting a highly obscured AGN.   {\it A. Constantin \& A. C. Seth}, present the peculiar nature of the nucleus of M94, which is the least luminous broad-line (type 1) LINER with possibly the least luminous broad line region known. 
 This is followed by an article from {\it R. DeCarli, R. Falomo, J.K. Kotilainen, T. Hyvonen, M. Uslenghi, \& A. Treves} on the extension of the scaling  relation between \SMBH  s and  host galaxies  to  the smallest observed local \SMBH s, showing that the relation holds over 2 dex in both M$_{\rm BH}$ and M$_{*}$.
Moreover, {\it N. Neumayer \& C. J. Walcher} and {\it P. Erwin \& D. A. Gadotti} extended the work to the nuclear star clusters and their embedded BHs, which are the possible precursors of massive black holes in galaxy nuclei.
 {\it N. Neumayer \& C. J. Walcher} studied the low mass end of the global-to-nucleus relations finding that the M$_{BH}$ - M$_{bulge}$ relation may well flatten at low masses while the M$_{\rm BH}$ - $\sigma$ relation may steepen.
Moreover, {\it P. Erwin \& D. A. Gadotti} analyze a sample of disk galaxies and performing a 2D bulge/disk/bar decompositions, show that while SMBHs correlate with the stellar mass of the bulge component of galaxies, the masses of  nuclear star clusters correlate much better with the total galaxy stellar mass.

We  conclude the Special Issue by discussing 
what we  can learn from studying  the large scale  environment of AGN.
{\it R.  Fassbender, R.   Suhada, \& A.  Nastasi} studied the distribution of X-ray AGN in the large-scale structure environments of 22 X-ray luminous galaxy clusters in the redshift range 0.9$<z<$1.6. They found two overdensities, one at r$<$1 Mpc of predominantly low-luminosity AGN and another one of brighter soft-band detected AGN at cluster-centric distances of 2-3 Mpc (about three times the average cluster radius R$_{200}$ of the systems).
Their results support the idea of two different physical triggering mechanisms of X-ray AGN activity in dependence of the radially changing large-scale structure environment of the distant clusters.
{\it M. Gitti, F. Brighenti,  \& B.R. McNamara} draw a qualitative picture  of the current knowledge of the effects of the AGN feedback on the Intra Cluster Medium by summarizing the recent results in this field. 
Finally, {\it N. Cappelluti, V. Allevato, \& A. Finoguenov} review the now abundant data
on AGN small and large scale clustering properties from different deep and large surveys
and the full important information we can derive from them.

\end{document}